**Room-temperature local ferromagnetism and nanoscale domain growth in the ferromagnetic semiconductor $Ge_{1-x}Fe_x$**

Yuki K. Wakabayashi,[1] Shoya Sakamoto,[2] Yukiharu Takeda,[3] Keisuke Ishigami,[2] Yukio Takahashi,[2] Yuji Saitoh,[3] Hiroshi Yamagami,[3] Atsushi Fujimori,[2] Masaaki Tanaka,[1] and Shinobu Ohya[1]

[1]*Department of Electrical Engineering and Information Systems, The University of Tokyo, 7-3-1 Hongo, Bunkyo-ku, Tokyo 113-8656, Japan*

[2]*Department of Physics, The University of Tokyo, Bunkyo-ku, Tokyo 113-0033, Japan*

[3]*Synchrotron Radiation Research Unit, JAEA, Sayo, Hyogo 679-5148, Japan*

Abstract

We investigate the local electronic structure and magnetic properties of the group-IV-based ferromagnetic semiconductor, $Ge_{1-x}Fe_x$ (GeFe), using soft X-ray magnetic circular dichroism. Our results show that the doped Fe $3d$ electrons are strongly hybridized with the Ge $4p$ states, and have an unusually large orbital magnetic moment relative to the spin magnetic moment; i.e., $m_{orb}/m_{spin} \approx 0.3$. We find that local ferromagnetic domains, which are formed through ferromagnetic exchange interactions in the high-Fe-content regions of the GeFe films, exist at room temperature, well above the Curie temperature of 20 - 100 K. We demonstrate the first observation of the intriguing nanoscale domain growth process in which ferromagnetic domains expand as the temperature decreases, followed by a transition of the entire film into a ferromagnetic state at the Curie temperature.

A major issue that must be addressed for the realization of semiconductor spintronic devices is to achieve room-temperature ferromagnetism in ferromagnetic semiconductors (FMSs) based on the widely used III-V and group-IV materials. In $Ga_{1-x}Mn_xAs$ (GaMnAs), which is a particularly well-studied FMS, the highest Curie temperature ($T_C$) ever reported is 200 K [1]. In GaMnAs, $T_C$ is limited by the presence of interstitial Mn atoms, which are antiferromagnetically coupled to the substitutional Mn atoms [2]. Recently, however, the group-IV-based FMS, $Ge_{1-x}Fe_x$ (GeFe), has been reported to exhibit several attractive features [3 – 5]. It can be grown epitaxially on Si and Ge substrates without the formation of intermetallic precipitates, and is therefore compatible with mature Si process technology. Unlike GaMnAs, with GeFe, interstitial Fe atoms do not lead to a decrease in $T_C$ [6], and $T_C$ can be easily increased to above 200 K by thermal annealing [7]. Furthermore, $T_C$ does not depend on the carrier concentration, and thus $T_C$ and resistivity can be controlled separately [8], which is a unique feature that is only observed with GeFe and is a considerable advantage in overcoming the conductivity mismatch problem between ferromagnetic metals and semiconductors in spin-injection devices. Despite these attractive features, a detailed microscopic understanding of the ferromagnetism in GeFe, which is vitally important for room-temperature applications, is lacking. Here, we investigate the local magnetic behavior of GeFe using X-ray magnetic circular dichroism (XMCD), which is a powerful technique for element-specific detection of magnetic moments [9 – 12]. We find that local ferromagnetic domains remain in the GeFe films even at room temperature, i.e., well above $T_C$; it follows that GeFe potentially has strong ferromagnetism, which persists even at room temperature. Furthermore, we present the first observations of the intriguing feature that ferromagnetic domains, which are formed above $T_C$ via the ferromagnetic exchange interaction in high-Fe concentration regions of the films, develop and grow as the temperature decreases, and that all of them coalesce at temperatures below $T_C$. Such a nanoscale domain growth process is a key feature in understanding materials that exhibit *homogeneous* ferromagnetism (i.e., where the film is free from any ferromagnetic precipitates) despite the *inhomogeneous* distribution of magnetic atoms in the film [6,7].

We carried out XMCD measurements on two samples (labeled A and B) consisting of a 2-nm-thick Ge capping layer, and a 120-nm-thick $Ge_{0.935}Fe_{0.065}$ layer with a 30-nm-thick Ge buffer layer grown on a Ge(001) substrate by low-temperature molecular beam epitaxy (LT-MBE) [Figs. 1(a) and 1(b)]. The Ge buffer and Ge cap layers were grown at 200°C, and the $Ge_{0.935}Fe_{0.065}$ layer of sample A was grown at 160°C, whereas that of sample B was grown at 240°C [6]. From the Arrott plots of the magnetic field ($H$) dependence of the magnetic circular dichroism (MCD) measured with visible light with a photon energy of 2.3 eV (corresponding to the $L$-point energy gap of bulk Ge), we found $T_C = 20$ K and 100 K for samples A and B, respectively. Detailed crystallographic analyses showed that the GeFe films are single crystalline, with a diamond-type structure and nanoscale spatial Fe concentration fluctuations of 4% − 7% (sample A) and 3% − 10% (sample B) [6]. We found that $T_C$ is higher when the fluctuations in the Fe concentration are larger. In addition, channeling Rutherford backscattering and channeling particle-induced X-ray emission measurements showed that ~85% (~15%) of the doped Fe atoms exist at the substitutional (tetrahedral interstitial) sites in both samples A and B, and that the interstitial Fe concentration is not related to $T_C$ [6]. This indicates that there are no ferromagnetic precipitates with different crystal structures in our films. We performed X-ray absorption spectroscopy (XAS) and XMCD measurements at the twin-helical undulator beamline BL23SU of SPring-8 [13]. The XAS spectra were obtained in total electron yield mode. To remove the oxidized surface layer, the samples were briefly etched in dilute hydrofluoric acid (HF) prior to loading into the XAS (XMCD) vacuum chamber.

We measured XAS spectra [$\mu^+$, $\mu^-$, and ($\mu^+ + \mu^-$)/2] at the $L_2$ (~721 eV) and $L_3$ (~708 eV) absorption edges of Fe in sample A [Fig. 1(c)] and B [Fig. 1(d)] at 5.6 K with $\mu_0 H = 5$ T applied perpendicular to the film surface. Here, $\mu^+$ and $\mu^-$ refer to the absorption coefficients for photon helicity parallel and antiparallel to the Fe $3d$ majority spin direction, respectively. In both films, the three peaks $a$, $b$, and $c$ are observed at the Fe $L_3$ edge in the XAS spectra [see also the insets in Figs. 1(c) and 1(d)]. We found that the small peak $c$ was suppressed by etching the surface with dilute HF, indicating that this peak, which can be assigned to the $Fe^{3+}$ state, originates from a small quantity of surface Fe oxide [14], which remains even after surface cleaning. Meanwhile, peaks $a$ and $b$ are assigned to the Fe atoms in GeFe [15,16].

We measured the XMCD (= $\mu^+ - \mu^-$) spectra at the Fe $L_2$ and $L_3$ absorption edges of samples A [Fig. 1(e)] and B [Fig. 1(f)] at 5.6 K with various $H$ applied perpendicular to the film surface. Here, we discuss the XMCD intensities at 707.66 eV (X) and 708.2 eV (Y), which correspond to the photon energies of peaks $a$ and $b$ in the XAS spectra, respectively. When normalized to 707.3 eV, the XMCD spectra with various $H$ differ, and the intensity at X grows faster than that at Y as $H$ increases, as shown in the insets of Figs. 1(e) and 1(f). As shown in Figs. 1(c) and (d), the shapes of the XAS spectra at the Fe $L_3$ edge are similar between samples A and B, which have almost the same interstitial Fe concentrations (i.e., 15% of the total Fe content [6]); therefore, we can assign the XMCD intensity at X to the substitutional Fe atoms and the paramagnetic component of the XMCD intensity at Y to the interstitial Fe atoms. We do not observe fine structures due to multiplet splitting at the Fe $L_3$ edge in both samples, which would be observed if the $3d$ electrons were localized and were not strongly hybridized with other orbitals [17]. These observations indicate that the Fe $3d$ electrons are strongly hybridized with the Ge $4p$ states [18].

We determine the orbital magnetic moment, $m_{orb}$, and the spin magnetic moment, $m_{spin}$, of the *substitutional* Fe atoms from the XAS and XMCD spectra at the $L_{2,3}$ edge region of Fe using the XMCD sum rules [19 − 23] [see Section I of Supplemental Material (SM)]. As shown in Figs. 2(a) and 2(b), both $m_{spin}$ and $m_{orb}$ (and therefore the total magnetic moment $M = m_{spin} + m_{orb}$) are larger in sample B ($T_C = 100$ K) than in sample A ($T_C = 20$ K) over the entire temperature region when $\mu_0 H = 5$ T. For sample A, $m_{orb}/m_{spin}$ = 0.31 ± 0.02, and for sample B, $m_{orb}/m_{spin}$ = 0.30 ± 0.03, both of which are positive and significantly larger than that of bulk Fe (where $m_{orb}/m_{spin} \sim 0.043$ [19]); the orbital angular momentum in GeFe is not quenched. The observation that the spin and orbital angular momentum are parallel suggests that the Fe $3d$ shell is more than half filled. This implies that the Fe atoms are in the $2^+$ state rather than in the $3^+$ state, in which the Fe $3d$ shell is half-filled and the orbital angular momentum vanishes. This large $m_{orb}$ is a characteristic property of GeFe, and excludes the possibility of the existence of ferromagnetic Fe metal precipitates in our films.

Figure 2(c) shows the $H$ dependence of the XMCD intensity at energy X and a temperature of 5.6 K (blue curve), the MCD intensity measured with visible light of 2.3 eV at 5 K (red dotted curve), and the magnetization measured using a superconducting quantum interference device (SQUID) at 5 K (green dotted curve) for sample B. The shapes of these curves show excellent agreement with each other; it follows that the spin splitting of the valence band composed of the Ge $4p$ orbitals is induced by the Fe $3d$ magnetic moment, which originates from the substitutional Fe atoms, through the $p$-$d$ hybridization. The lower panels of Fig. 2 show the effective magnetic-field ($H_{eff}$) dependence of the XMCD intensity measured at X for samples A (d) and B (e) at various temperatures. Here, $M$ is also plotted (filled red symbols), and $\mu_0 H_{eff}$ is obtained by subtracting the product of $M$ and the sheet density of the substitutional Fe atoms from $\mu_0 H$ to eliminate the influence of the demagnetization field. The scale of the vertical axis of the XMCD intensity is adjusted so that it represents $M$ at each temperature. The insets show clear hysteresis below $T_C$ in both samples. The XMCD $- H_{eff}$ curves show large curvature above $T_C$ in both samples [see the main panels of Figs. 2(d) and 2(e)], indicating that part of the film is superparamagnetic (SPM) above $T_C$. It indicates that ferromagnetic domains form in nanoscale high-Fe concentration regions at temperatures above $T_C$, and thus $M$ can be described by

$$M = 5.2 f_{SPM} L(\frac{m_{SPM}\mu_0 H_{eff}}{k_B T}) + (1\text{-}f_{SPM})\frac{C}{T}\mu_0 H_{eff},$$ (1)

where $f_{SPM}$ and $m_{SPM}$ are fitting parameters expressing the fraction of substitutional Fe atoms which participate in the SPM component, and the magnetic moment per ferromagnetic domain, respectively. Also, $C$ is the Curie constant per substitutional Fe atom (see Section III of the SM), and $L$ is the Langevin function. Here, 5.2 is the *ideal* saturated value of $M$; i.e., $M = m_{spin} + (m_{orb}/m_{spin}) \times m_{spin}$, where $m_{spin} = 4$ $\mu_B$ (for $Fe^{2+}$) and $m_{orb}/m_{spin} \approx 0.3$ [Figs. 2(a) and 2(b)] when all the substitutional Fe atoms are magnetically active. The first and second terms in Eq. (1) correspond to the SPM and paramagnetic components, respectively. In Figs. 2(d) and 2(e), the black dashed curves correspond to the best fit obtained with Eq. (1). For sample B, the $M - H_{eff}$ curves at temperatures in the range $100 - 300$ K are well reproduced by Eq. (1), which indicates that the ferromagnetic $-$ SPM transition occurs at $T_C = 100$ K. With sample A, the $M - H_{eff}$ curves at temperatures above $T_C$ (i.e., $T > 20$ K) are well reproduced by Eq. (1), except for $T = 20$ K, which is probably due to the onset of ferromagnetism. These good fits up to room temperature indicate that ferromagnetic interactions within the nanoscale high-Fe concentration regions remain at room temperature in both samples.

The residual $M$, which is obtained from a linear extrapolation of $M$ from the high magnetic field region to $H_{eff} = 0$ at 5.6 K, is 1.2 $\mu_B$ per Fe atom in sample A, and 1.5 $\mu_B$ per Fe atom in sample B. This result suggests that only ~23% (= 1.2/5.2) and ~29% (= 1.5/5.2) of the substitutional Fe atoms are magnetically active in samples A and B, respectively. In Figs. 2(d) and 2(e), the high-field magnetic susceptibilities $\partial M/ \partial(\mu_0 H_{eff})$ ($\mu_B$/T per Fe atom) at 4 T and 5.6 K are 0.15 in sample A and 0.10 in sample B. Because $\partial M/\partial(\mu_0 H_{eff})$ at 4 T per substitutional paramagnetic Fe atom should be 0.37 (see Section IV of the SM), this result indicates that the ratios of paramagnetic Fe atoms to the total number of Fe atoms are only ~41% (= 0.15/0.37) and ~27% (= 0.10/0.37), respectively. This means that some fraction of the moment of the Fe atoms is missing, and thus suggests that there are Fe atoms that couple antiferromagnetically with the ferromagnetic Fe atoms in the films. This is also supported by the weak spin-glass behavior observed in GeFe at very low temperatures [7].

We see a similar trend in the temperature dependence of the fitting parameters $f_{SPM}$ and $m_{SPM}$ in both films; i.e., $f_{SPM}$ and $m_{SPM}$ both increase with decreasing temperature (Fig. 3). This result means that the ferromagnetic domains, which form only in the nanoscale high-Fe concentration regions at room temperature [Fig. 4(a)], expand toward lower Fe concentration regions with decreasing temperature [Fig. 4(b)], and finally the entire film becomes ferromagnetic at $T_C$ [Fig. 4(c)]. This appears to be a characteristic feature of materials that exhibit *homogeneous* ferromagnetism, despite the *inhomogeneous* distribution of magnetic atoms in the film [6,7]. As shown in Fig. 3, $f_{SPM}$ and $m_{SPM}$ are larger in sample B than in sample A, which is attributed to the difference in spatial fluctuations of the Fe concentration, which are 4% − 7% in sample A and 3% − 10% in sample B [6]. The larger the nonuniformity of the Fe distribution is, the larger ferromagnetic domains, $f_{SPM}$, and $m_{SPM}$ become, and the domains can more easily connect magnetically, resulting in a higher $T_C$.

In summary, we have investigated the local electronic structure and magnetic properties of the doped Fe atoms in the $Ge_{0.935}Fe_{0.065}$ films using XAS and XMCD. The Fe atoms appear in the $2^+$ state, with the $3d$ electrons strongly hybridized with the $4p$ electrons in Ge; this results in a delocalized $3d$ nature and long-range ferromagnetic ordering, leading to the excellent agreement between the $H$ dependence of magnetization, MCD, and XMCD. Using the XMCD sum rules, we obtained the $M − H_{eff}$ curves, which can be explained by the coexistence of SPM and paramagnetic ordering at temperatures above $T_C$. The fitting results clearly show that the local ferromagnetic domains, which exist at room temperature, expand with decreasing temperature, leading to a ferromagnetic transition of the entire system at $T_C$. The nonuniformity of the Fe concentration plays a crucial role for the formation of the magnetic domains, and our results indicate that strong ferromagnetism is inherent to GeFe, and persists at room temperature.

**ACKNOWLEDGEMENTS**

We would like to thank T. Okane for support with the experiments. This work was partly supported by Grants-in-Aid for Scientific Research (22224005, 23000010, and 26249039) including the Specially Promoted Research, and the Project for Developing Innovation Systems from MEXT. This work was performed under the Shared Use Program of JAEA Facilities (Proposal No. 2014A-E31) with the approval of the Nanotechnology Platform Project supported by MEXT. The synchrotron radiation experiments were performed at the JAEA beamline BL23SU in SPring-8 (Proposal No. 2014A3881). Y. K. Wakabayashi and Y. Takahashi acknowledge financial support from JSPS through the Program for Leading Graduate Schools (MERIT). S. Sakamoto acknowledges financial support from JSPS through the Program for Leading Graduate Schools (ALPS).

# References

[1] L. Chen, X. Yang, F. Yang, J. Zhao, J. Misuraca, P. Xiong, and S. von Molnár, *Nano Lett.* **11**, 2584 (2011).

[2] K. M. Yu, W. Walukiewicz, T. Wojtowicz, I. Kuryliszyn, X. Liu, Y. Sasaki, and J. K. Furdyna, Phys. Rev. B **65**, 201303 (2002).

[3] Y. Shuto, M. Tanaka, and S. Sugahara, phys. stat. sol. **3**, 4110 (2006).

[4] Y. Shuto, M. Tanaka, and S, Sugahara, Appl. Phys. Lett. **90**, 132512 (2007).

[5] Y. Shuto, M. Tanaka, and S. Sugahara, Jpn. J. Appl. Phys. **47**, 7108 (2008).

[6] Y. K. Wakabayashi, S. Ohya, Y. Ban, and M. Tanaka, J. Appl. Phys. **116**, 173906 (2014).

[7] Y. K. Wakabayashi, Y. Ban, S. Ohya and M. Tanaka, Phys. Rev. B **90**, 205209 (2014).

[8] Y. Ban, Y. Wakabayashi, R. Akiyama, R. Nakane, and M. Tanaka, AIP Advances **4**, 097108 (2014).

[9] D. J. Keavney, D. Wu, J. W. Freeland, E. J. Halperin, D. D. Awschalom, and J. Shi, Phys. Rev. Lett. **91**, 187203 (2003).

[10] K. W. Edmonds, N. R. S. Farley, T. K. Johal, G. V. D. Laan, R. P. Campiom, B. L. Gallegher, and C. T. Foxon, Phys. Rev. B **71**, 064418 (2005).

[11] D. J. Keavney, S. H. Cheung, S. T. King, M. Weinert, and L. Li, Phys. Rev. Lett. **95**, 257201 (2005).

[12] V. R. Singh, K. Ishigami, V. K. Verma, G. Shibata, Y. Yamazaki, T. Kataoka, A. Fujimori, F. -H. Chang, D. -J. Lin, C. T. Chen, Y. Yamada, T. Fukumura, and M. Kawasaki, Appl. Phys. Lett. **100**, 242404 (2012).

[13] Y. Saitoh, Y. Fukuda, Y. Takeda, H. Yamagami, S. Takahashi, Y. Asano, T. Hara, K. Shirasawa, M. Takeuchi, T. Tanaka, and H. Kitamura, J. Synchrotron Rad. **19**, 388 (2012)

[14] T. J. Regan, H. Ohldag, C. Stamm, F. Nolting, J. Luning, J. Stohr, and R. L. White, Phys. Rev. B **64**, 214422 (2001).

[15] R. Kumar, A. P. Singh, P. Thakur, K. H. Chae, W. K. Choi, B. Angadi, S. D. Kaushik, and S. Panaik, J. Phys. D: Appl. Phys. **41**, 155002 (2008).

[16] E. Sakai, K. Amemiya, A. Chikamatsu, Y. Hirose, T. Shimada, and T. Hasegawa, J. Magn. Magn. Mater, **333**, 130 (2013).

[17] G. V. D. Laan and I. W. Kirkman, J. Phys.: Condens. Matter **4**, 4189 (1992).

[18] I. A. Kowalik, A. Persson, M. A. Nino, A. Na, A. Navarro-Quezada, B. Faina, A. Bonanni, T. Dietl, and D. Arvanits, Phys. Rev. B **85**, 184411 (2012).

[19] C. T. Chen, Y. U. Idzerda, H. -J. Lin, N. V. Smith, G. Meigs, E. Chaban, G. H. Ho, E. Pellegrin, and F. Sette, Phys. Rev. Lett. **75**, 152 (1995).

[20] Y. Takeda, M. Kobayashi, T. Okane, T. Ohkochi, J. Okamoto, Y. Saitoh, K. Kobayashi, H. Yamagami, A. Fujimori, A. Tanaka, J. Okabayashi, M. Oshima, S. Ohya, P. N. Hai, and M. Tanaka, Phys. Rev. Lett. **100**, 247202 (2008).

[21] K. Mamiya, T. Koide, A. Fujimori, H. Tokano, H. Manaka, A. Tanaka, H. Toyosaki, T. Fukumura, and M. Kawasaki, Appl. Phys. Lett. **89**, 062506 (2006).

[22] J. Stohr and H. Konig, Phys. Rev. Lett. **75**, 3748 (1995).

[23] C. Piamonteze, P. Miedema, and F. M. F. de Groot, Phys. Rev. B **80**, 184410 (2009).

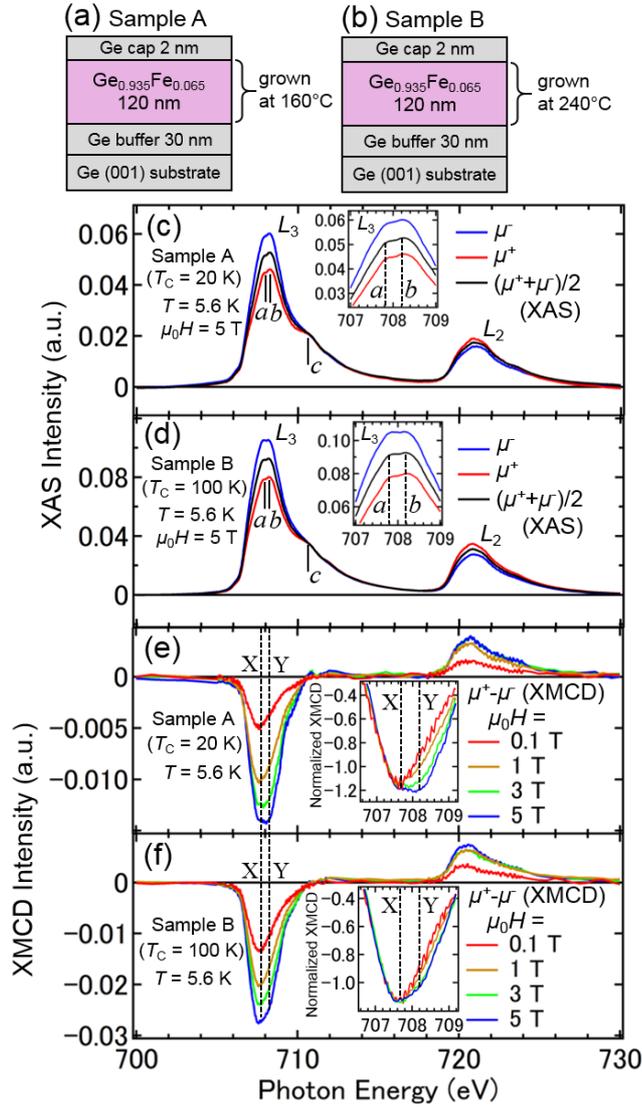

Fig. 1. (Color) Schematic diagrams showing the structures of sample A (a) and sample B (b). (c), (d) XAS spectra of $\mu^-$ (blue curve), $\mu^+$ (red curve), and $(\mu^+ + \mu^-)/2$ (black curve) at the $L_2$ (~721 eV) and $L_3$ (~708 eV) absorption edges of Fe for sample A (c) and sample B (d). The measurements were made with a magnetic field of $\mu_0 H = 5$ T applied perpendicular to the film surface at a temperature of 5.6 K. The insets show a magnified plot of the spectra at the Fe $L_3$ edge. (e), (f) XMCD ($= \mu^+ - \mu^-$) spectra at the $L_2$ and $L_3$ absorption edges of Fe for sample A (e) and sample B (f) measured at 5.6 K with $\mu_0 H = 0.1$ T (red curve), 1 T (brown curve), 3 T (green curve), and 5 T (blue curve) applied perpendicular to the film surface. The insets show a magnified plot of the spectra at the Fe $L_3$ edge, in which the XMCD data are normalized to 707.3 eV.

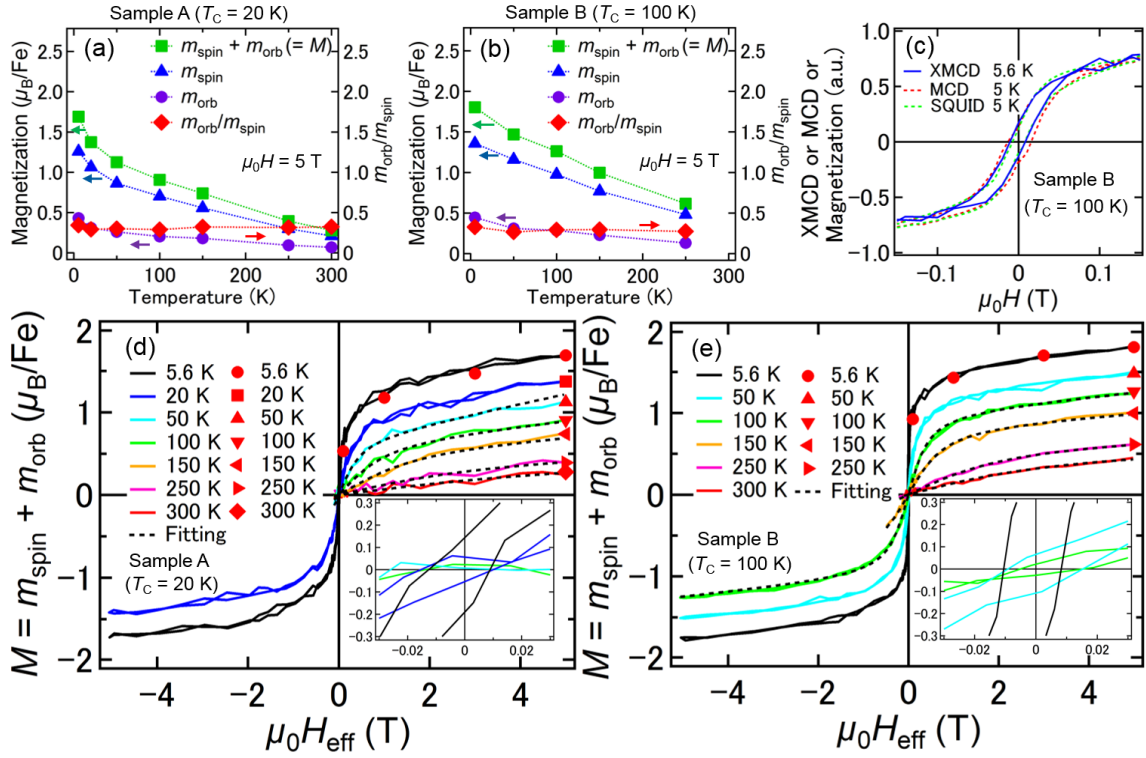

Fig. 2. (Color) (a), (b) The temperature dependence of $m_{spin} + m_{orb}$ (green squares), $m_{spin}$ (blue triangles), $m_{orb}$ (violet circles), and $m_{orb}/m_{spin}$ (red rhombuses) for sample A (a) and sample B (b) with an applied magnetic field of $\mu_0 H = 5$ T. (c) The $H$ dependence of the XMCD intensity (blue solid curve) at X shown in Fig. 1 (707.66 eV) at 5.6 K, the MCD intensity at 5 K with a photon energy of 2.3 eV corresponds to the $L$-point energy gap of bulk Ge (red dotted curve), and the magnetization measured using a SQUID at 5 K (green dotted curve) for sample B. (d),(e) The dependence of the XMCD intensity measured at X on the effective magnetic field $H_{eff}$ for sample A (d) and sample B (e) at various temperatures. The total magnetization ($M = m_{spin} + m_{orb}$) obtained using the XMCD sum rules is also plotted (filled red symbols). We scaled the vertical axis of the XMCD intensity so that it represents $M$ at each temperature. In all measurements, $H$ was applied perpendicular to the film surface.

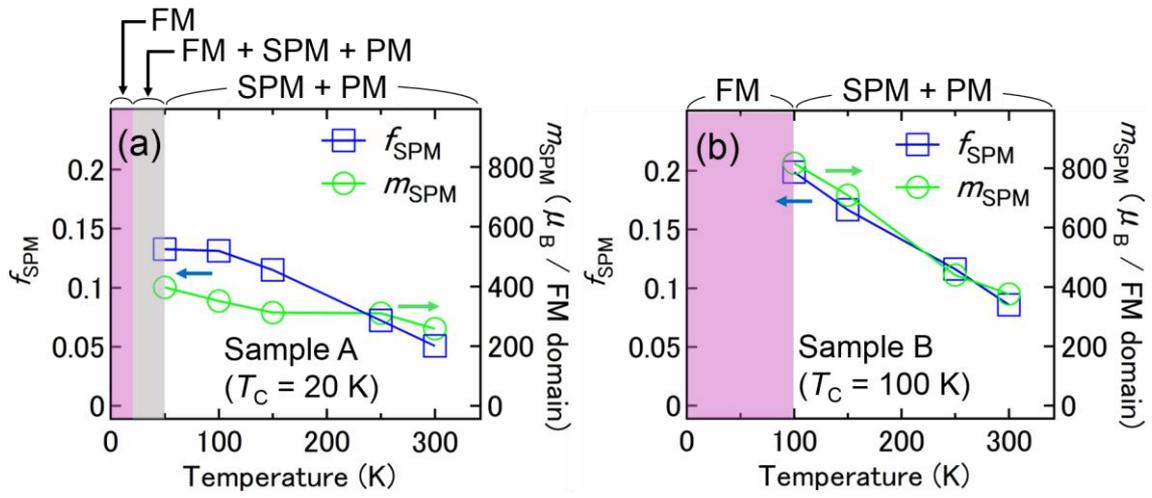

Fig. 3. (Color online) (a), (b) The temperature dependence of the best-fit parameters $f_{SPM}$ and $m_{SPM}$ obtained for sample A (a) and sample B (b). The red (or dark gray), gray, and white areas indicate ferromagnetic (FM), FM + SPM + paramagnetic (PM), and SPM + PM regions, respectively.

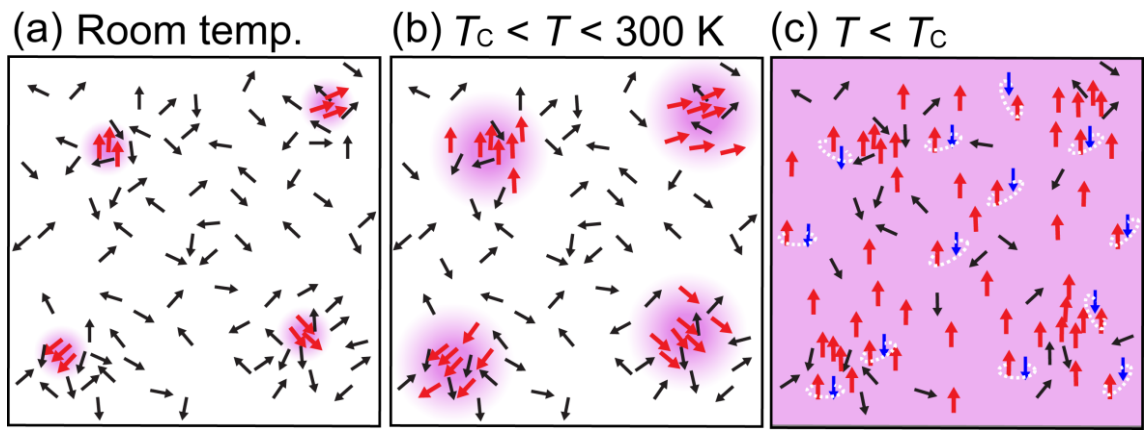

Fig. 4. (Color) (a) – (c) Schematic diagrams showing the magnetic states in the $Ge_{0.935}Fe_{0.065}$ films with zero magnetic field at room temperature (i.e., $T = 300$ K) (a), $T_C < T < 300$ K (b), and $T < T_C$ (c). The small black, red, and blue arrows correspond to the magnetic moments of the paramagnetic, ferromagnetic, and antiferromagnetically coupled substitutional Fe atoms, respectively. The red areas indicate ferromagnetic regions. Antiferromagnetically coupled Fe atoms are thought to exist all over the film at temperatures below $T_C$.

**Supplemental Material for Room-temperature local ferromagnetism and nanoscale domain growth in the ferromagnetic semiconductor Ge$_{1-x}$Fe$_x$**

Yuki K. Wakabayashi,[1] Shoya Sakamoto,[2] Yukiharu Takeda,[3] Keisuke Ishigami,[2] Yukio Takahashi,[2] Yuji Saitoh,[3] Hiroshi Yamagami,[3] Atsushi Fujimori,[2] Masaaki Tanaka,[1] and Shinobu Ohya[1]

[1]*Department of Electrical Engineering and Information Systems, The University of Tokyo, 7-3-1 Hongo, Bunkyo-ku, Tokyo 113-8656, Japan*
[2]*Department of Physics, The University of Tokyo, Bunkyo-ku, Tokyo 113-0033, Japan*
[3]*Synchrotron Radiation Research Unit, JAEA, Sayo, Hyogo 679-5148, Japan*

## I. Estimation of the spin magnetic moment and the orbital magnetic moment of substitutional Fe atoms using the X-ray magnetic-circular-dichroism sum rules

We obtain the spin magnetic moment $m_{spin}$ and the orbital magnetic moment $m_{orb}$ from the spectra of the X-ray absorption spectroscopy (XAS) and X-ray magnetic circular dichroism (XMCD) in the energy region near the $L_2$ and $L_3$ absorption edges of Fe using the XMCD sum rules.[S3]    Figure S1(a) shows the XAS spectra (solid curves) and the XAS signals integrated from 690 eV (dashed curves) of sample A.    Figure S1(b) shows the XMCD spectra (solid curves) and the XMCD signals integrated from 690 eV (dashed curves) of sample A.    Here, the measurements were carried out with a magnetic field $\mu_0 H = 5$ T applied perpendicular to the film surface at 5.6 K (black curves), 20 K (blue curves), 50 K (light blue curves), 100 K (green curves), 150 K (orange curves), 250 K (pink curves), and 300 K (red curves).    Figure S2 shows the same data measured for sample B.    For the XMCD sum-rules analyses, we define the values of $r$, $p$, and $q$ as the following equations at each temperature.

$$r = \int_{E_3+E_2} \frac{(\mu^+ + \mu^-)}{2} dE, \tag{S1}$$

$$p = \int_{E_3} (\mu^+ - \mu^-) dE, \tag{S2}$$

$$q = \int_{E_3+E_2} (\mu^+ - \mu^-) dE, \tag{S3}$$

where $E_3$ (690-718 eV) and $E_2$ (718-760 eV) represent the integration energy ranges for the $L_3$ and $L_2$ absorption edges, respectively.    Here, $\mu^+$ ($\mu^-$) and $E$ represent the absorption coefficient for the photon helicity parallel (antiparallel) to the Fe $3d$ majority spin direction and the incident photon energy, respectively. We can obtain $m_{\mathrm{spin}}$ and $m_{\mathrm{orb}}$ of substitutional Fe atoms using the XMCD sum rules, which are expressed as follows:

$$m_{\mathrm{orb}} = -\frac{2q}{3r}(10 - n_{3d}), \qquad (\mathrm{S4})$$

$$m_{\mathrm{spin}} + 7m_{\mathrm{T}} = -\frac{3p-2q}{r}(10 - n_{3d}), \qquad (\mathrm{S5})$$

where $n_{3d}$ and $m_{\mathrm{T}}$ are the number of $3d$ electrons on the Fe atom and the expectation value of the intra-atomic magnetic dipole operator, respectively. Because the paramagnetic component observed at Y in Figs. 1(e) and 1(f) in the main text, which originates from the interstitial Fe atoms, is negligibly small (See Section II of the Supplemental Material), the integrated values of the XMCD spectra $p$ [Eq. (S2)] and $q$ [Eq. (S3)] can be attributed only to the substitutional Fe atoms. Meanwhile, because the XAS signals have both contributions of the substitutional and interstitial Fe atoms, we reduced the integrated XAS intensity $r$ [Eq. (S1)] to 85% of its raw value (85% is the approximate ratio of the substitutional Fe atoms to that of the total Fe atoms in both samples A and B[S4]) when using the XMCD sum rules. We neglect the expectation value of the intra-atomic magnetic dipole operator, because it is negligibly small for Fe atoms at the $T_{\mathrm{d}}$ symmetry site.[S5] Also, we assume $n_{3d}$ to be 6 and the correction factor for the $m_{\mathrm{spin}}$ to be 0.88 for $Fe^{2+}$.[S6]

By the above calculations using the temperature dependence of XAS and XMCD spectra shown in Figs. S1 and S2, we obtained the temperature dependence of $m_{\mathrm{spin}}$ and $m_{\mathrm{orb}}$ of substitutional Fe atoms as shown in Figs. 2(a) and 2(b) in the main text.

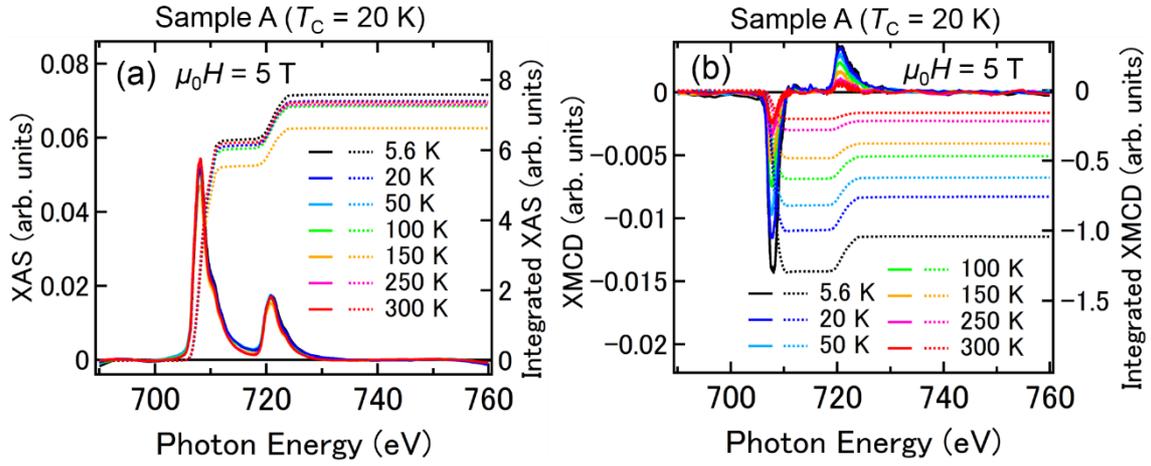

Fig. S1. (a) XAS [= $(\mu^+ + \mu^-)/2$] spectra (solid curves) and the XAS signals integrated from 690 eV (dashed curves) of sample A. (b) XMCD (= $\mu^+ - \mu^-$) spectra (solid curves) and the XMCD signals integrated from 690 eV (dashed curves) of sample A. These measurements were carried out with a magnetic field $\mu_0 H = 5$ T applied perpendicular to the film surface at 5.6 K (black curves), 20 K (blue curves), 50 K (light blue curves), 100 K (green curves), 150 K (orange curves), 250 K (pink curves), and 300 K (red curves).

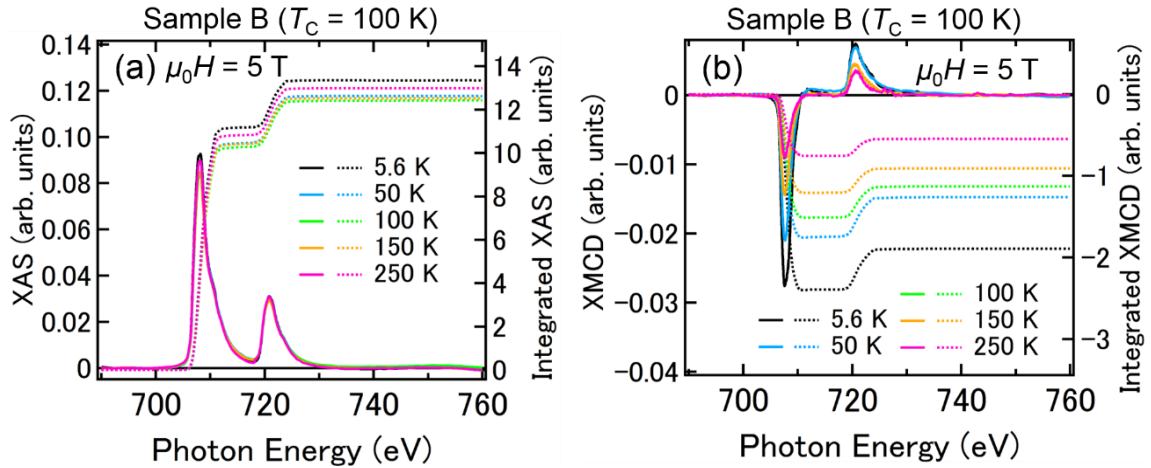

Fig. S2. (a) XAS [= $(\mu^+ + \mu^-)/2$] spectra (solid curves) and the XAS signals integrated from 690 eV (dashed curves) of sample B. (b) XMCD (= $\mu^+ - \mu^-$) spectra (solid curves) and the XMCD signals integrated from 690 eV (dashed curves) of sample B. These measurements were carried out with a magnetic field $\mu_0 H = 5$ T applied perpendicular

to the film surface at 5.6 K (black curves), 50 K (light blue curves), 100 K (green curves), 150 K (orange curves), and 250 K (pink curves).

## II. Influence of the paramagnetic XMCD component on the XMCD sum-rules analyses

Figures S3 shows the XMCD spectra of samples A (a) and B (b) normalized to 707.3 eV measured at 5.6 and 300 K with magnetic fields of 0.1 and 5 T applied perpendicular to the film surface. In both films, all the line shapes of the XMCD spectra are almost the same, which means that the paramagnetic component observed at Y in Figs. 1(e) and 1(f) of the main text is negligibly small for the XMCD sum-rules analyses and that the integrated XMCD signal can be attributed only to the substitutional Fe atoms.

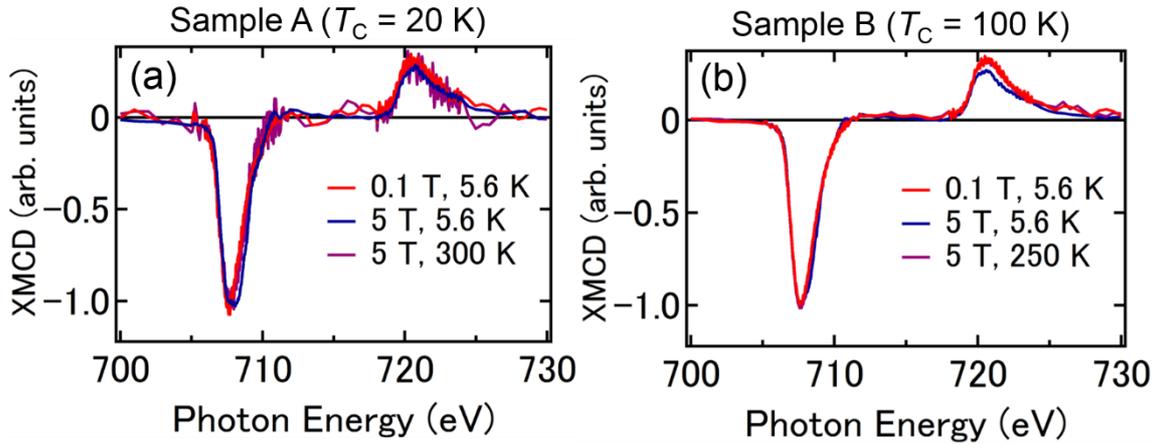

Fig. S3. XMCD spectra of samples A (a) and B (b) normalized to 707.3 eV measured at 5.6 and 300 K with magnetic fields of 0.1 and 5 T applied perpendicular to the film surface.

## III. Estimation of the Curie constant per substitutional Fe atom

The Curie constant per substitutional Fe atom is obtained using the following equations:

$$C = \frac{\mu_B^2}{3k_B} n_B^2,$$ (S6)

$$n_B = \left[\frac{3}{2} + \frac{S(S+1) - L(L+1)}{2J(J+1)}\right]\sqrt{J(J+1)}, \tag{S7}$$

where $\mu_B$, $k_B$, $n_B$, $S$, $L$, and $J$ represent the Bohr magneton, the Boltzmann constant, the effective Bohr magneton number, the spin angular momentum, the orbital angular momentum, and the total angular momentum, respectively. We obtained $n_B$ assuming $S = 2$ (for $Fe^{2+}$), $L = 1.2$ ($L = 2S \times m_{orb}/m_{spin}$, where $m_{orb}/m_{spin} \approx 0.3$ as shown in Figs. 2(a) and 2(b) in the main text), and $J = 3.2$ ($= L + S$ because the spin and orbital angular momenta of a substitutional Fe atom are parallel) in Eq. (S7). Thus, $n_B$ is estimated to be 5.96.

## IV.   Estimation of the high-field magnetic susceptibility of a substitutional paramagnetic Fe atom

At very low temperature below ~20 K, the effective magnetic-field $H_{eff}$ dependence of the total magnetization $M$ ($= m_{spin} + m_{orb}$) of one substitutional paramagnetic Fe atom is expressed by the Langevin function. Thus, the $H_{eff}$ dependence of $M$ of one substitutional paramagnetic Fe atom at 5.6 K is obtained by substituting 5.2 $\mu_B$, 1, and 5.6 K in $m_{SPM}$, $f_{SPM}$, and $T$ of Eq. (1) in the main text, respectively (Fig. S4). We approximated the high-field magnetic susceptibility $\partial M/\partial(\mu_0 H_{eff})$ ($\mu_B$/T per Fe) at 4 T by the slope of the $M$-$H_{eff}$ line from 4 T to 5 T (black dashed line in Fig. S4). In this way, $\partial M/\partial(\mu_0 H_{eff})$ at 4 T is estimated to be 0.37 $\mu_B$/T per one substitutional paramagnetic Fe. In Figs. 2(d) and 2(e) in the main text, the $\partial M/\partial(\mu_0 H_{eff})$ values at 4 T and 5.6 K in samples A and B are 0.15 and 0.10, respectively; it follows that the ratios of paramagnetic Fe atoms to the total number of Fe atoms are ~41% ($= 0.15/0.37$) in sample A and ~27% ($= 0.10/0.37$) in sample B.

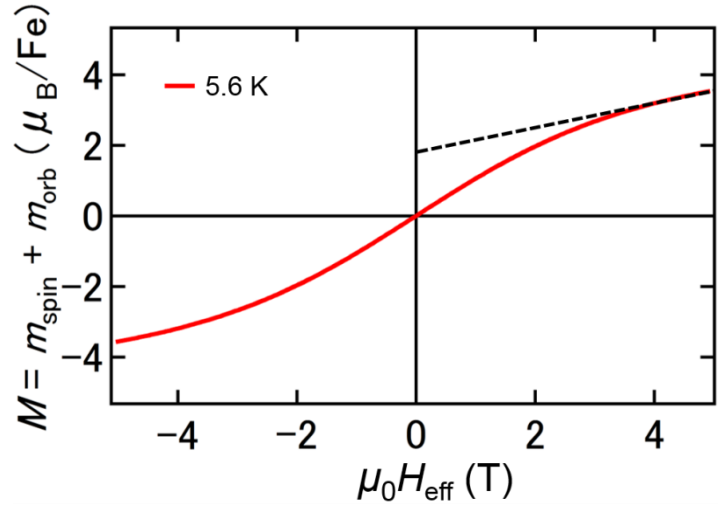

Fig. S4. Effective magnetic-field $H_{\text{eff}}$ dependence of the total magnetization $M$ (= $m_{\text{spin}}+m_{\text{orb}}$) per one substitutional paramagnetic Fe at 5.6 K obtained using Eq. (1) in the main text.

**References**

[S3] C. T. Chen, Y. U. Idzerda, H. -J. Lin, N. V. Smith, G. Meigs, E. Chaban, G. H. Ho, E. Pellegrin, and F. Sette, Phys. Rev. Lett. **75**, 152 (1995).

[S4] Y. K. Wakabayashi, S. Ohya, Y. Ban, and M. Tanaka, J. Appl. Phys. **116**, 173906 (2014).

[S5] J. Stohr and H. Konig, Phys. Rev. Lett. **75**, 3748 (1995).

[S6] C. Piamonteze, P. Miedema, and F. M. F. de Groot, Phys. Rev. B **80**, 184410 (2009).